\documentclass[a4paper,11pt]{article}
\usepackage{pos}
\usepackage[export]{adjustbox}
\usepackage{caption}

\newlength{\bibitemsep}
\newlength{\bibparskip}
\let\oldthebibliography\thebibliography
\renewcommand\thebibliography[1]{%
  \oldthebibliography{#1}%
  \setlength{\parskip}{\bibitemsep}%
  \setlength{\itemsep}{\bibparskip}%
}

\title{ Prompt atmospheric leptons and 
   the potential role of 
    intrinsic charm}

\author*[a]{Laksha Pradip Das}
\author[a]{Diksha Garg}
\author[b]{Maria Vittoria Garzelli}
\author[a]{Mary Hall Reno}
\author[b]{G\"unter Sigl}

\affiliation[a]{Department of Physics and Astronomy,
University of Iowa,\\
  Iowa City, IA,  USA}
\affiliation[b]{II Institut fuer Theoretische Physik, 
Universitaet Hamburg,\\
Luruper Chaussee 149, D-22761 Hamburg, Germany}

\emailAdd{lakshapradip-das@uiowa.edu}
\emailAdd{diksha-garg@uiowa.edu}
\emailAdd{maria.vittoria.garzelli@desy.de}
\emailAdd{mary-hall-reno@uiowa.edu}
\emailAdd{guenter.sigl@desy.de}

\abstract{The  all-sky very-high energy ($10^4-10^6$ GeV) atmospheric muon flux is most recently measured by IceCube, where in the higher energy range, the spectrum hardens indicating a prompt component.
IceCube also measures the atmospheric muon neutrino flux at high energy. Since this is dominated by the astrophysical flux, they are only able to set an upper bound on the prompt atmospheric muon neutrino  flux contribution. We provide a new evaluation of the prompt
atmospheric muon flux including for the first time an intrinsic charm component to colliding nucleons.  This increases forward production of  $\bar{D}^{0}$, $D^-$ and $\Lambda_c$ which decay into final states that can contain muons and muon neutrinos.
We show how the increase in the prompt muon flux due to intrinsic charm has an associated increase in the prompt muon neutrino flux.  
We consider the Regge ansatz for intrinsic charm production that we implement in MCEq used for the calculation of the lepton fluxes. We discuss the challenges of obtaining predictions that are simultaneously consistent with both IceCube's high energy atmospheric muon flux measurements and their upper bound on the prompt muon neutrino flux. We quantify the discrepancies.}

\ConferenceLogo{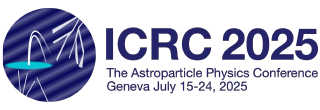}

\FullConference{39th International Cosmic Ray Conference (ICRC2025)\\
 15–24 July 2025\\
Geneva, Switzerland\\}

\begin{document}
\maketitle

\section{Introduction}

The measurement of the flux of atmospheric neutrinos played an important role in our understanding of neutrino flavor oscillations \cite{Super-Kamiokande:1998kpq}. The atmospheric muon and neutrino fluxes are related, as they come from the decays of hadrons produced by cosmic ray interactions with air nuclei. The so-called conventional flux comes from the decays of pions and kaons that are copiously produced. Cosmic ray production of heavy flavor, notably charm hadrons, followed by their rapid decays in semileptonic channels contribute to the prompt atmospheric lepton fluxes.
Prompt muon and muon neutrino fluxes from charm decay are equal to within $\sim 15-20\%$ . As emphasized in ref. \cite{Illana:2010gh}, another component of the prompt atmospheric muon flux is given by high energy light unflavored mesons with prompt electromagnetic decays and branching fractions of order $\sim 10^{-4}$ to final states that include a $\mu^+\mu^-$ pair,
e.g., from $B(\eta\to \mu^+\mu^-\gamma)=(3.1\pm 0.4 )\times 10^{-4}$ \cite{ParticleDataGroup:2024cfk}. Light unflavored mesons do not contribute to the high energy prompt atmospheric neutrino flux.
The flux of prompt atmospheric neutrinos is isotropic for neutrino energies below $\sim 10^7$ GeV,
nominally a background to the diffuse astrophysical neutrino flux which is also isotropic. 
The flux of prompt atmospheric muons is also isotropic.

\begin{figure}[h]
    \centering
    \includegraphics[width=0.475\linewidth]{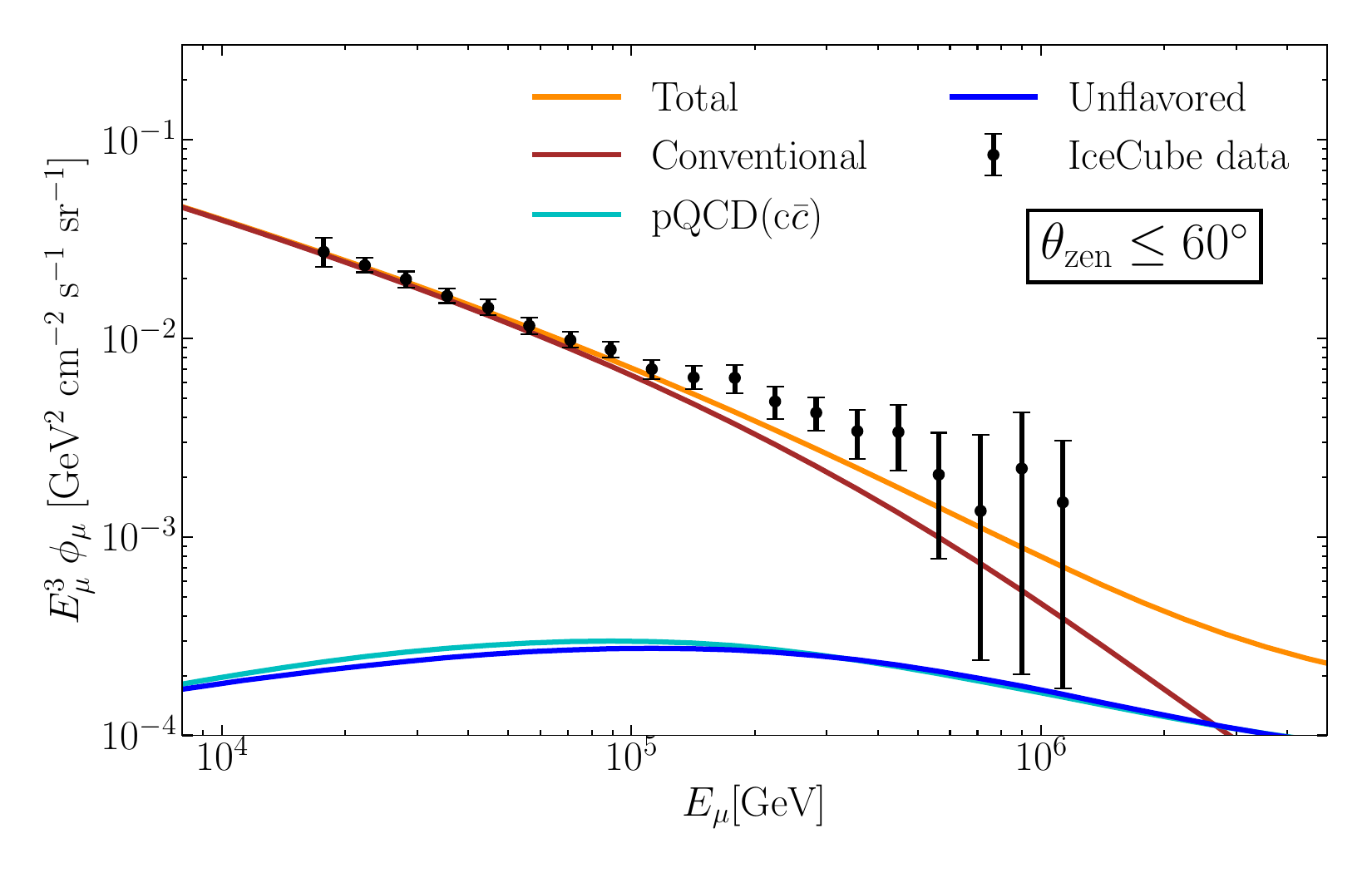}
        \includegraphics[width=0.475\linewidth]{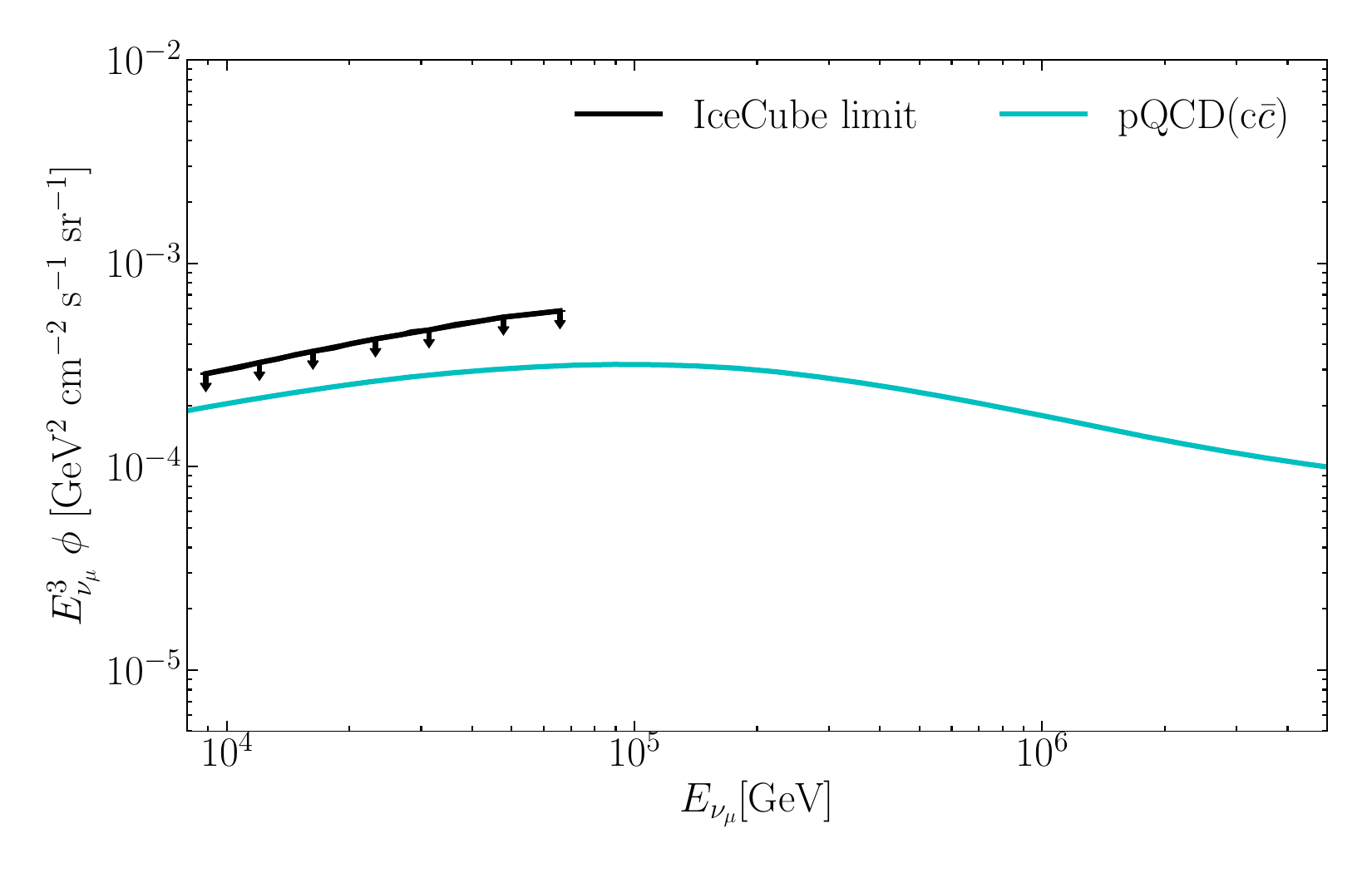}
    \caption{Angle-averaged muon flux ($\mu^{+}+\mu^{-}$) for $\theta_{\rm zen}<60^\circ$ from MCEq vs. IceCube measured spectrum of high-energy muons \cite{Soldin:2023lbr,Soldin:2018vak,IceCube:2015wro} with bin-by-bin error bars (left). All-sky averaged prompt ($\nu_\mu + \bar{\nu}_{\mu}$) spectrum from charm vs. IceCube upper limit on prompt neutrino flux \cite{IceCube:2016umi} determined from up-going muon track measurements, {assuming that the dependence of the prompt neutrino flux on energy is the same as the energy dependence of the Enberg et al. (ERS) prompt $\nu_\mu + \bar{\nu}_{\mu}$\cite{Enberg:2008te}  flux prediction} (right). Both fluxes are computed with MCEq using as input the cosmic ray all-nucleon spectrum H3a \cite{Gaisser:2011klf} and hadronic interaction model  SIBYLL-2.3c \cite{Riehn:2017mfm,Fedynitch:2018cbl}. Note the different $y$-axis scales in the two panels.}
    \label{fig:average-flux-muon-neutrino}
\end{figure}

The prompt atmospheric neutrino flux has not yet been detected \cite{IceCube:2016umi}. On the other hand, the atmospheric muon flux has been measured by the IceCube collaboration at energies higher than $E_\mu=1$
PeV \cite{Soldin:2023lbr,Soldin:2018vak,IceCube:2015wro}. 
The measured atmospheric muon flux shown in the left panel of fig. \ref{fig:average-flux-muon-neutrino} indicates a potential spectral break that may point to a shift from muons produced by pion and kaon decays to muons produced by prompt decays of heavy flavor and unflavored mesons. 
The data plotted here are angle-averaged for zenith angles $\theta<60^\circ$  \cite{IceCube:2015wro}, zenith angles that avoid subtleties associated with nearly horizontal muons, both in their production and propagation, and geometrical effects due to the curvature of the Earth. In this work, we use the MCEq framework \cite{Fedynitch:2018cbl,fedynitch2015calculationconventionalpromptlepton} for theoretical predictions, with the so-called H3a model of cosmic-ray composition and all-nucleon energy spectrum\cite{Gaisser:2011klf} as implemented in MCEq and hadronic interaction model SIBYLL-2.3c \cite{Riehn:2017mfm,Fedynitch:2018cbl}, also available as default in MCEq. Charm production in MCEq is implemented in SIBYLL-2.3c with a phenomenological model based on soft and hard minijets, as described in detail in ref. \cite{Fedynitch:2018cbl}. A comparison of SIBYLL-2.3c results with LHCb data and with the perturbative results of refs. \cite{Gauld:2015yia,Garzelli:2016xmx,PROSA:2015yid} show a reasonable agreement \cite{Fedynitch:2018cbl}. 
In what follows, we denote MCEq's implementation of charm production as "standard" perturbative QCD (pQCD($c\bar{c}$)) to distinguish it from other charm production mechanisms, in particular, from intrinsic charm.
We have implemented $\Lambda_c$ decays in our modification of the default MCEq software \cite{Das:2025tbd}.
The orange curve in the left panel of Fig.~\ref{fig:average-flux-muon-neutrino} sums the conventional and prompt contributions to the atmospheric muon flux, where the unflavored and pQCD charm contributions are nearly equal. The right panel of fig. \ref{fig:average-flux-muon-neutrino} shows the all-sky averaged IceCube upper bound on the prompt neutrino flux \cite{IceCube:2016umi} 
and the MCEq prediction. 

From the energy spectrum, the IceCube collaboration provides a best-fit prompt component (charm and unflavored combined) for the atmospheric muon flux as 4.75 units of the Enberg et al. (ERS) prompt atmospheric muon flux from charm, assumed equal to the prompt atmospheric muon neutrino flux \cite{Enberg:2008te}. The ERS prediction is consistent with the MCEq results for standard perturbative charm contributions to the atmospheric lepton fluxes. Nominally, MCEq perturbative charm plus unflavored meson contributions to the prompt atmospheric muon flux shown in the left panel of fig.~\ref{fig:average-flux-muon-neutrino} account for less than half of IceCube's best-fit prompt atmospheric muon flux.
Direct constraints on intrinsic charm contributions to the atmospheric neutrino flux have been derived (see, e.g., refs. \cite{Garzelli:2023jlq,Laha:2016dri,Halzen:2016thi}). 
Given hints of ``missing'' prompt components in the atmospheric muon flux, we consider whether or not enhanced open charm production from intrinsic charm can account for the deficit in the atmospheric muon flux prediction. We {fit the normalization  of the intrinsic charm contribution to the prompt atmospheric muon flux measured in IceCube \cite{IceCube:2015wro}, then use the same normalization to evaluate the intrinsic charm contribution to} the prompt neutrino flux.
This is similar to approaches that use comparisons of predictions to measurements of the  conventional muon flux to tune models and thereby reduce uncertainties in the associated predictions of the conventional neutrino flux ~\cite{Honda:2006qj,Yanez:2023lsy}.

\section{Intrinsic charm model}
\label{sec:IC}

\begin{figure}
    \centering
\includegraphics[width=0.4\linewidth,valign=c]{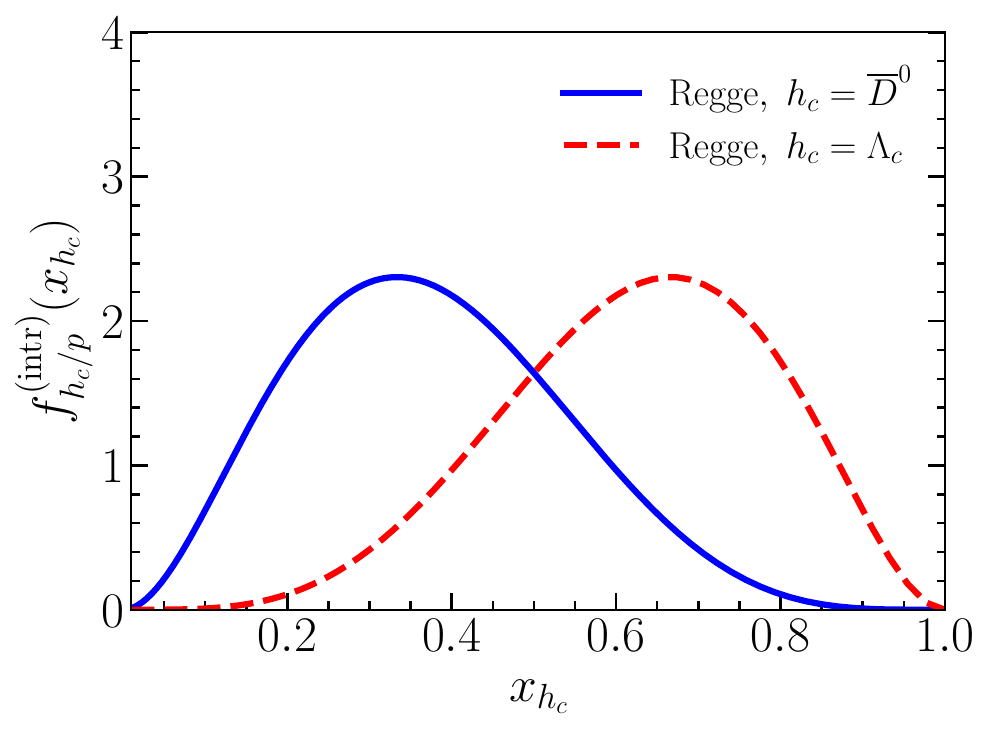}
    \hfil
    \begin{minipage}{\dimexpr 0.5\linewidth-\columnsep}
        \captionsetup{singlelinecheck=off, skip=0pt}
      \caption{Fragmentation functions $f_{h_c/p}^{\rm (intr)} (x)$ for $h_c=\bar D^0, \Lambda_c$ and $x_{h_c}=E_{h_c}/E_p$ following the Regge ansatz \cite{Kaidalov:1985jg,Kaidalov:2003wp}, 
    {normalized to unity}.}
    \label{fig:fragmentation}
    \end{minipage}
    \vspace{-2ex}
\end{figure}

We implement intrinsic charm with a normalization factor $w_{\rm intr}^c $, not predicted by first principles, that multiplies the proton-air inelastic cross section and a fragmentation function $f_{h_c}^{\rm intr}$ for the proton to fluctuate into the charm hadron $h_c$ following ref. \cite{Ostapchenko:2022thy,Garzelli:2023jlq}.  Explicitly, we take
\begin{equation}
  \frac{d\sigma_{p-{\rm air}}^{h_c{\rm (intr)}} (E,x_{h_c})}{dx_{h_c}} = w_{\rm intr}^c \sigma _{p-{\rm air}}(E) f_{h_c/p}^{\rm (intr)} (x_{h_c})\,,
  \label{eq:IC}
\end{equation}
{where $x_{h_c}=E_{h_c}/E_p$
for proton energy $E_p$ incident on the air nucleus and outgoing charm hadron energy $E_{h_c}$.} 
The charm hadrons $h_c=\bar D^0,\, D^-$ and $\Lambda_c$ are produced according to
\begin{eqnarray}
    pA &\to \bar D^0\Lambda_c X\\
     nA &\to  D^-\Lambda_c X \,.\,
\end{eqnarray}
for charm production plus any final state $X$. The Regge ansatz \cite{Kaidalov:1985jg,Kaidalov:2003wp} {for the meson fragmentation function (normalized to unity) is}
\begin{eqnarray}
f_{{h_c}/p}^{\rm (intr)} (x)&=& N_{\rm intr}\, x^{-a_\psi}(1-x)^{-a_
\psi + 2(1-a_N)}
    \label{eq:Regge}\\
  N_{\rm intr}   &=& \frac{\Gamma (-2a_\psi+4-2a_N)}{\Gamma(-a_\psi+1)\Gamma(-a_\psi+3-2a_N)}\,.
  \nonumber
\end{eqnarray}
{We use $a_\psi = -2.0$ and $a_N = -0.5$ as in ref. \cite{Garzelli:2023jlq}.}

We assume isospin symmetry for neutron-air interactions, so $\sigma_{p-{\rm air}}=\sigma_{n-{\rm air}}$. 
We assume that $f_{D^-/n}^{\rm (intr)}(x_D) = f_{\bar D^0/p}^{\rm (intr)}(x_D)$ and 
$f_{\Lambda_c/p}^{\rm (intr)}(x_\Lambda)
    = f_{\bar D^0/p}^{\rm (intr)}(1-x_\Lambda)= f_{\Lambda_c/n}^{\rm (intr)}(x_\Lambda)$.
We assume that a formula analogous to eq. (\ref{eq:IC}) applies also to neutron-air interactions.
Figure \ref{fig:fragmentation} shows the fragmentation functions for $\bar D^0$ and $\Lambda_c$. {The differential distribution in eq. (\ref{eq:IC}) times the cosmic ray all-nucleon flux is integrated to yield the prediction for charm hadron production via intrinsic charm. Because the hadrons produced via intrinsic charm take such a large fraction of the incident nucleon energy, the leptons in their decays can make large contributions to the prompt atmospheric lepton fluxes, even with very small values of $w_{\rm intr}^c$.}

\section{Theory predictions vs. IceCube experimental data}

\begin{figure}
    \centering
    \includegraphics[width=0.475\linewidth]{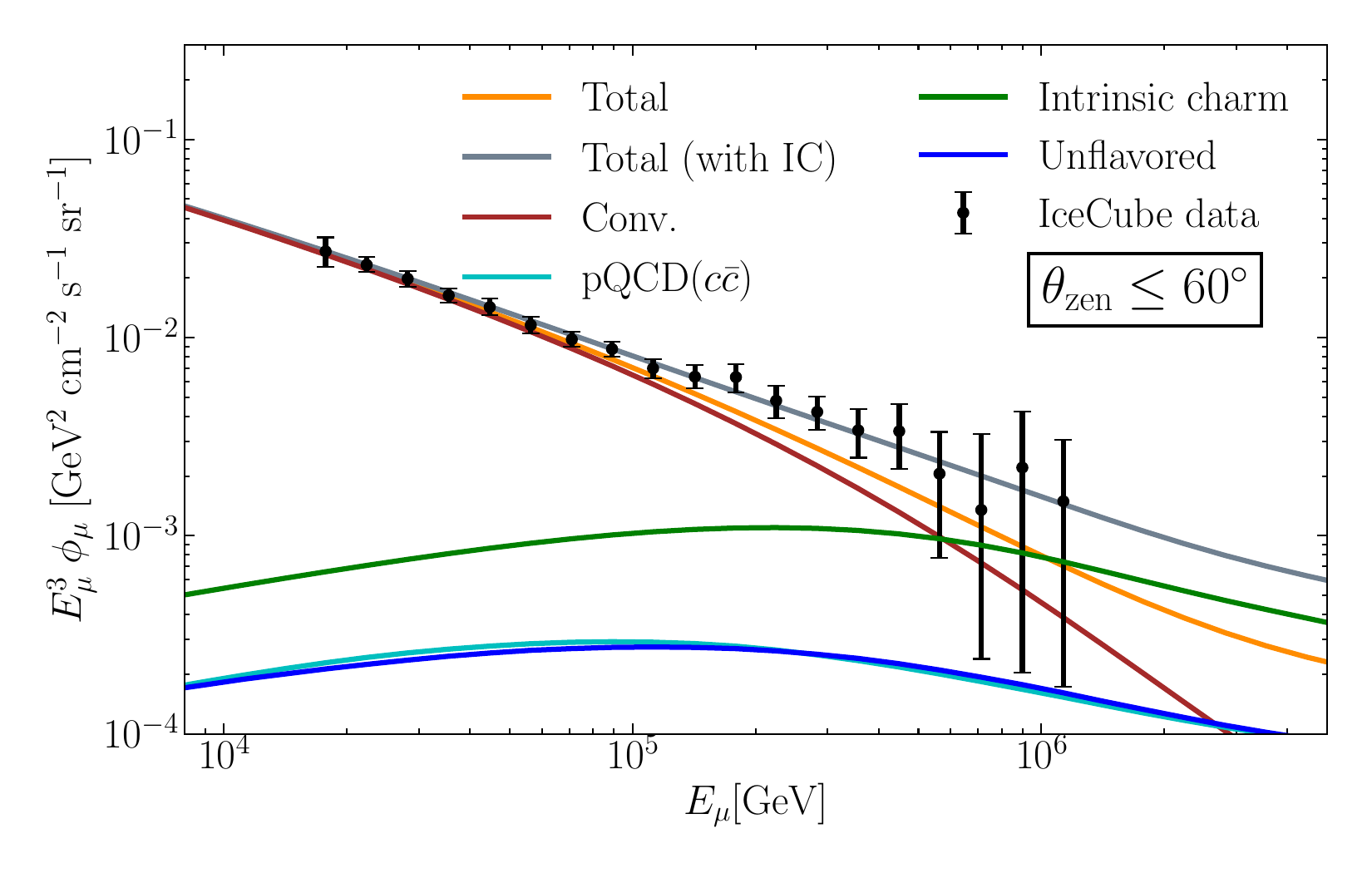}
        \includegraphics[width= 0.475\linewidth]{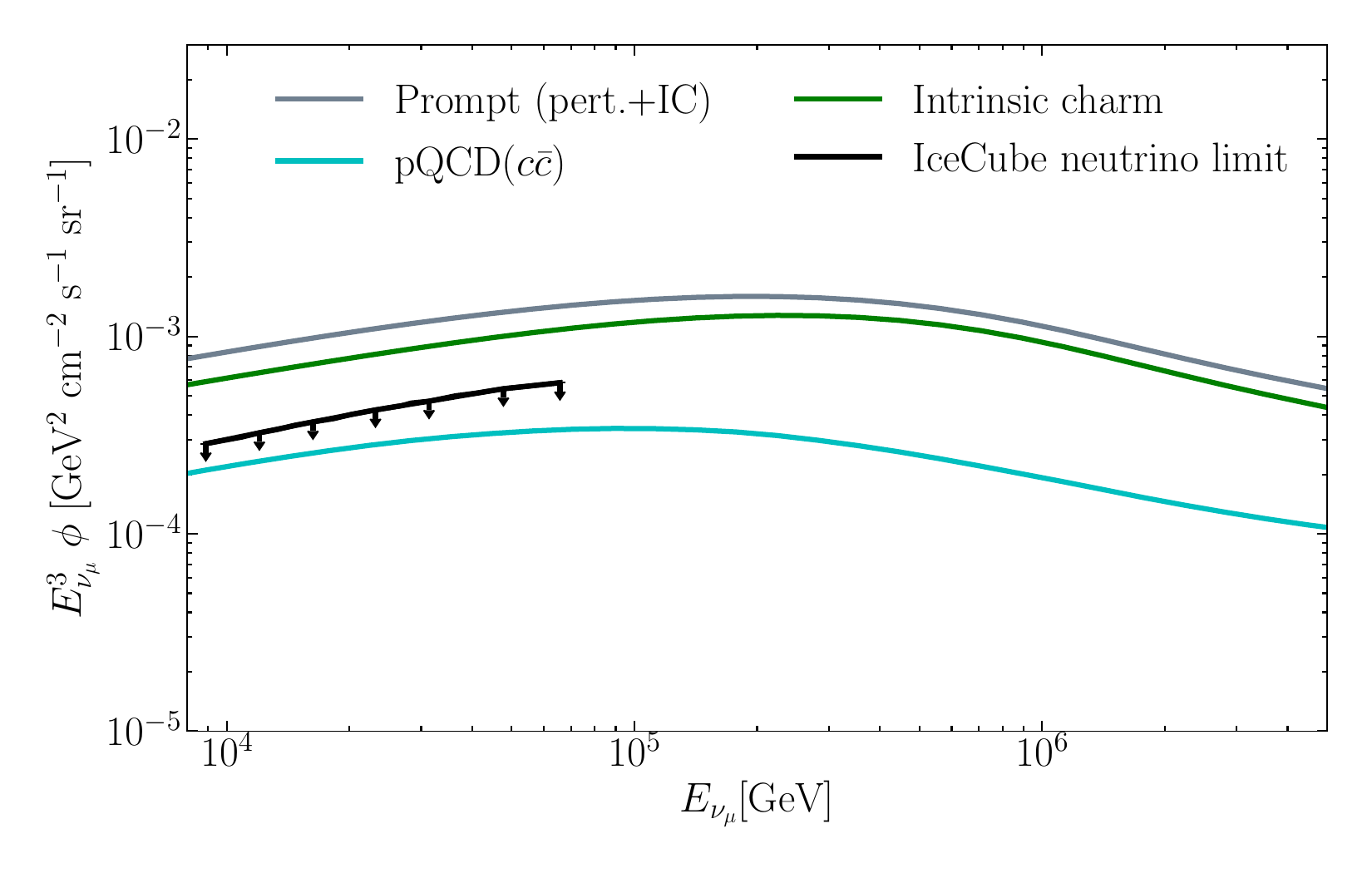}
    \caption{Same as in fig. \ref{fig:average-flux-muon-neutrino}, but with the addition of an intrinsic charm contribution with weight $w^c_{\rm intr}=0.0039$.  
    }
    \label{fig:best_fit_muons}
\end{figure}

We begin by including intrinsic charm contributions to the prompt atmospheric muon flux. With our assumptions outlined in sec. \ref{sec:IC}, the only parameter that is varied is 
$w_{\rm intr}^c$. 
The best-fit value
{to the angle-averaged muon flux for $\theta_{\rm zen} = 0^\circ-60^\circ$ using IceCube data \cite{IceCube:2015wro} including the total uncorrelated bin-by-bin uncertainties} is $w_{\rm intr}^c=0.0039$ $\pm \,0.001$ 
using the minimum $\chi^2$ estimation method.
The predicted atmospheric muon flux including intrinsic charm (IC) is shown in the left panel  of fig. \ref{fig:best_fit_muons} for zenith angle $\theta_{\rm zen}<60^\circ$.
Such an amount of intrinsic charm over-predicts the prompt atmospheric muon neutrino flux, as shown in the right panel of fig. \ref{fig:best_fit_muons}.

\begin{figure}
    \centering    \includegraphics[width=0.475\linewidth]{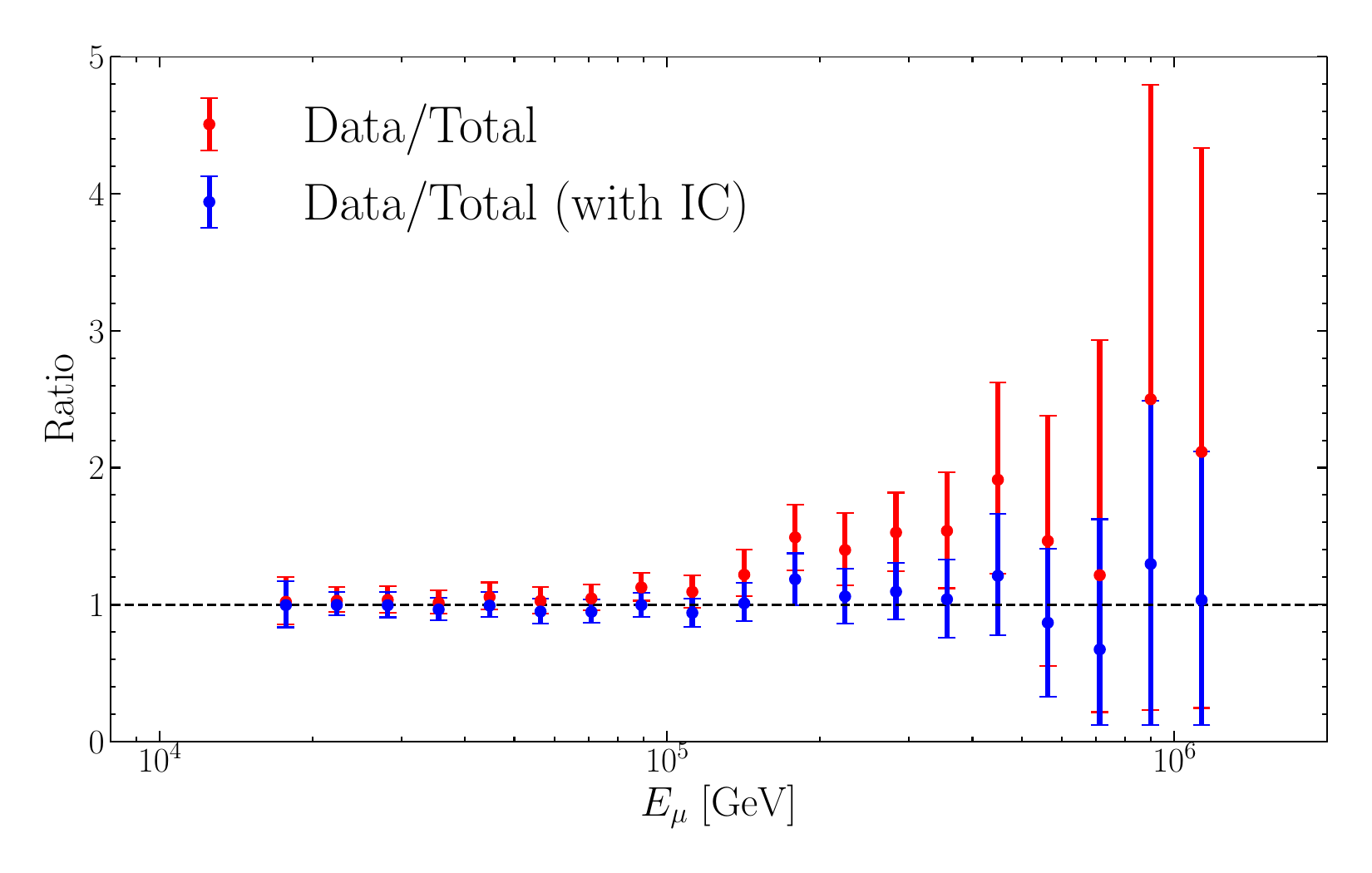} \includegraphics[width=0.475\linewidth]{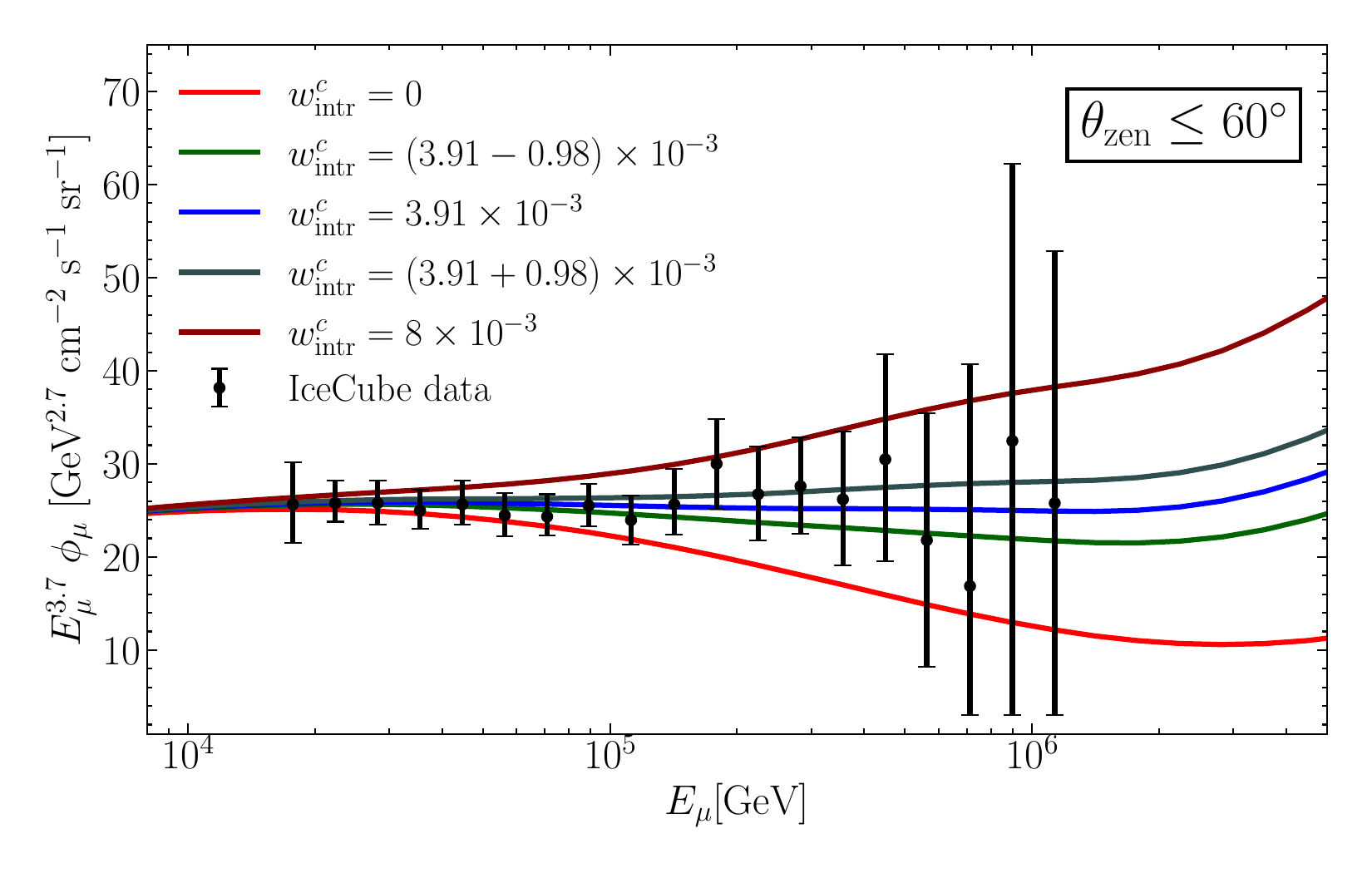}
    \caption{Ratio of IceCube muon data \cite{Soldin:2023lbr,Soldin:2018vak,IceCube:2015wro} to the muon flux from MCEq with ($w^c_{\rm intr}  = 0.0039$) and without including intrinsic charm contribution (left).  Total atmospheric muon flux with different $w^c_{\rm intr}$ (right). In these figures, $\alpha=1.$}
    \label{fig:Ratio}
\end{figure}

To gauge the effects of including intrinsic charm with $w_{\rm intr}^c=0.0039$, fig. \ref{fig:Ratio} compares the data to MCEq predictions with and without intrinsic charm. The left panel shows the ratio of the muon data to the muon flux from MCEq, both angle averaged for zenith angle $\theta_{\rm zen}<60^\circ$, including intrinsic charm with 
$w_{\rm intr}^c=0.0039$ (blue) and without intrinsic charm (red). The right panel shows the IceCube data for $E_\mu^{3.7}\phi_\mu$ averaged over zenith angles $\theta_{\rm zen}<60^\circ$ together with curves corresponding to theory predictions obtained with $w_{\rm intr}^c=0$ (red), {0.0039 (best fit, in blue), 0.0039 $\pm$ 0.001 (gray and green, respectively) and 0.008 (brown).}

\begin{figure}
    \centering
    \includegraphics[width=0.475\linewidth]{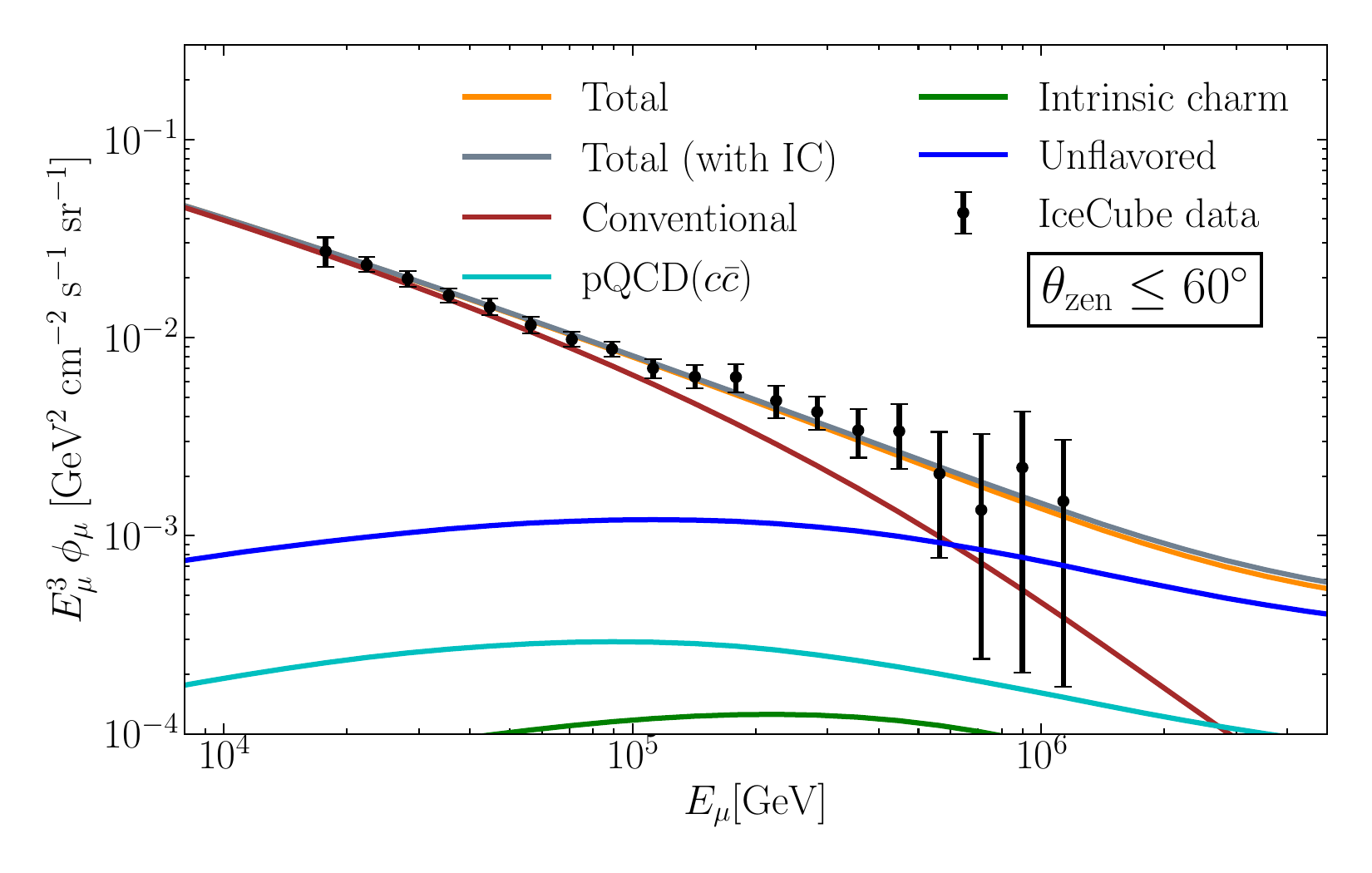}
    \includegraphics[width=0.475\linewidth]{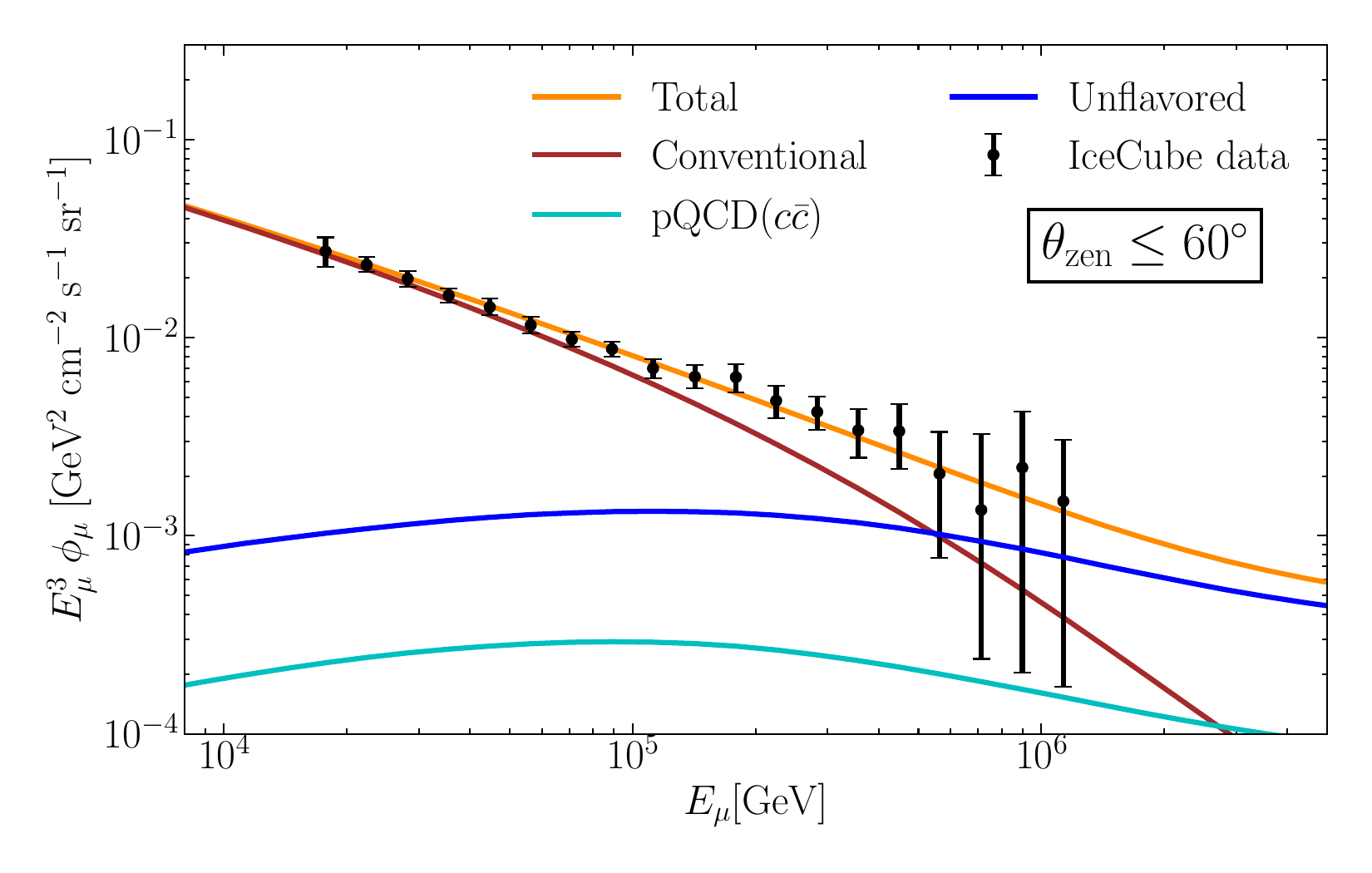}
    \caption{Same 
    as in the left panel of fig. \ref{fig:average-flux-muon-neutrino}, but including an intrinsic charm contribution with $w^c_{\rm intr}  = 4.46\times 10^{-4}$ and scaling the unflavored flux by a best-fit parameter of $\alpha = 4.39$ (left), or, alternatively, without intrinsic charm and 
    with unflavored contribution scaled by
    $\alpha = 4.84$ (right). 
    }
    \label{fig:Muon_flux_unfl_scaled}
\end{figure}

An alternative approach to set an upper limit to the intrinsic charm contribution is to fix $w_{\rm intr}^c$ to the maximum value permitted by the upper bound on the prompt atmospheric muon neutrino plus antineutrino flux. This yields a value of $w^c_{\rm intr} =4.46\times 10^{-4}$, significantly smaller than what is required for the best fit to the atmospheric muon data. 
{Unflavored mesons also contribute to the prompt atmospheric muon flux with a nearly identical energy dependence as the contribution from pQCD charm production and decay. We compensate for the nominal deficit in the theory prediction of prompt atmospheric muon flux by rescaling the unflavored contribution by $\alpha$, which we assume to be a constant.} 
Given $w^c_{\rm intr} =4.46\times 10^{-4}$, the left panel of fig. \ref{fig:Muon_flux_unfl_scaled} shows the prediction for $\alpha = 4.39$.  For reference, the right panel of fig. \ref{fig:Muon_flux_unfl_scaled} includes only a rescaling of the unflavored contribution without including any intrinsic charm. A factor of  $\alpha=4.84$ times the unflavored prompt contribution to the atmospheric muon flux 
{is the best fit without an intrinsic charm contribution. }
The values of $\alpha$ are so similar because for $w^c_{\rm intr} =4.46\times 10^{-4}$, intrinsic charm (green curve in left panel of fig. \ref{fig:Muon_flux_unfl_scaled}) provides a very small contribution to the atmospheric muon flux at high energies.

\section{Discussion}
\label{sec:discussion}

Accounting for intrinsic charm to achieve nominally better agreement between theory predictions and the IceCube measurements of the high-energy atmospheric muon flux yields a prediction for the all-sky prompt atmospheric $\nu_\mu+\bar\nu_\mu$ flux that exceeds the IceCube upper limit by a factor of $\sim 2.7$ at $E_{\nu_\mu}=10^4$ GeV. At the same energy, the upper bound on the all-sky prompt atmospheric $\nu_\mu+\bar\nu_\mu$ flux is a factor of $\sim 1.4$ above the MCEq pQCD prediction, as shown in fig. \ref{fig:best_fit_muons}. 
If, instead, one uses the upper bound on the all-sky prompt atmospheric $\nu_\mu+\bar\nu_\mu$ flux, intrinsic charm has a minor contribution to the atmospheric muon flux, as shown in fig. \ref{fig:Muon_flux_unfl_scaled}. To achieve the nominal agreement to the degree exhibited in fig. \ref{fig:Ratio}, an enhancement factor of $\alpha=4.39$ for the unflavored meson contribution is however required. 
In summary, taking our fits and comparisons at face value and assuming that both the IceCube analyses here considered are correct, intrinsic charm cannot account for the enhanced prompt contribution to the atmospheric muon flux at high energies, because it would violate the IceCube upper limit on prompt neutrino flux. On the other hand, even if the intrinsic charm contribution to $\nu_\mu+\bar\nu_\mu$ saturates the IceCube bound, the intrinsic charm contribution to the prompt muon flux is insufficient to reproduce {the measured atmospheric muon flux.} 
Work is in progress to better disentangle the signals of intrinsic charm, even if it does not dominate the prompt flux \cite{Das:2025tbd}.

On the other hand, if one neglects intrinsic charm and assumes that the discrepancy between data and the default MCEq atmospheric muon flux prediction is solely due to unflavored meson decays to muons, the unflavored contribution must be enhanced by a factor of $\alpha= 4.84$. 
The prompt muon fluxes from pQCD($c\bar c$) and from unflavored meson decays from MCEq with SIBYLL-2.3c are nearly equal for the energy range considered here (see left panel of fig. \ref{fig:average-flux-muon-neutrino}). With this model, we can define a function, $f_{\mu,\rm prompt}\equiv \phi_{\mu,{\rm pQCD}(c\bar{c})}\simeq \phi_{\mu,\rm unfl}$. 
Together, the best fit to the prompt contribution to the muon data is $\sim 4.84\,\phi_{\mu,\rm unfl}+\phi_{\mu,{\rm pQCD}(c\bar{c})}\simeq 5.84\, f_{\mu,\rm prompt}$. 
Theoretical uncertainties in pQCD charm pair production arising from missing higher orders are consistent with the possibility of a considerably larger pQCD contribution than the MCEq SIBYLL-2.3c result~\cite{Enberg:2008te,PROSA:2015yid,Gauld:2015yia,Bhattacharya:2015jpa,Bhattacharya:2016jce,Garzelli:2016xmx}.
However, even in this case, the upper bound on the prompt 
$\nu_\mu+\bar\nu_\mu$ flux $\phi_{\nu_\mu,{\rm pQCD}(c\bar{c})}$
constrains the extent  to which  
pQCD($c\bar c$) can contribute to the $\mu^++\mu^-$ flux. 
The IceCube prompt neutrino upper bound is approximately $\sim 1.4\, \phi_{\nu_\mu,{\rm pQCD}(c\bar{c})}$ \cite{IceCube:2016umi}.
This contribution amounts to $\sim 1.4 f_{\mu,{\rm prompt}}$, since $\phi_{\nu_\mu,{\rm pQCD}(c\bar{c})}\simeq \phi_{\mu,{\rm pQCD}(c\bar{c})}$ \cite{Lipari:2013taa}. The flux of muons from unflavored meson decays needed on top of the prompt perturbative contribution to saturate the experimental muon flux would still be quite high, $\sim 4.44 f_{\mu,\rm prompt}$ in this case. Other hadronic interaction models predict comparable (\texttt{DPMJET-III} and \texttt{EPOSLHC}) or lower (\texttt{QGSJET-II} and \texttt{QGSJET-III}) unflavored meson contributions to $\mu^++\mu^-$ than SIBYLL-2.3c. A large enhancement of forward unflavored meson production in proton-air collisions at high energies would be required in the modeling if unflavored meson decays are primarily responsible for the experimenally reported high-energy $\mu^++\mu^-$ flux.

\begin{figure}
    \centering
\includegraphics[width=0.5\linewidth,valign=c]{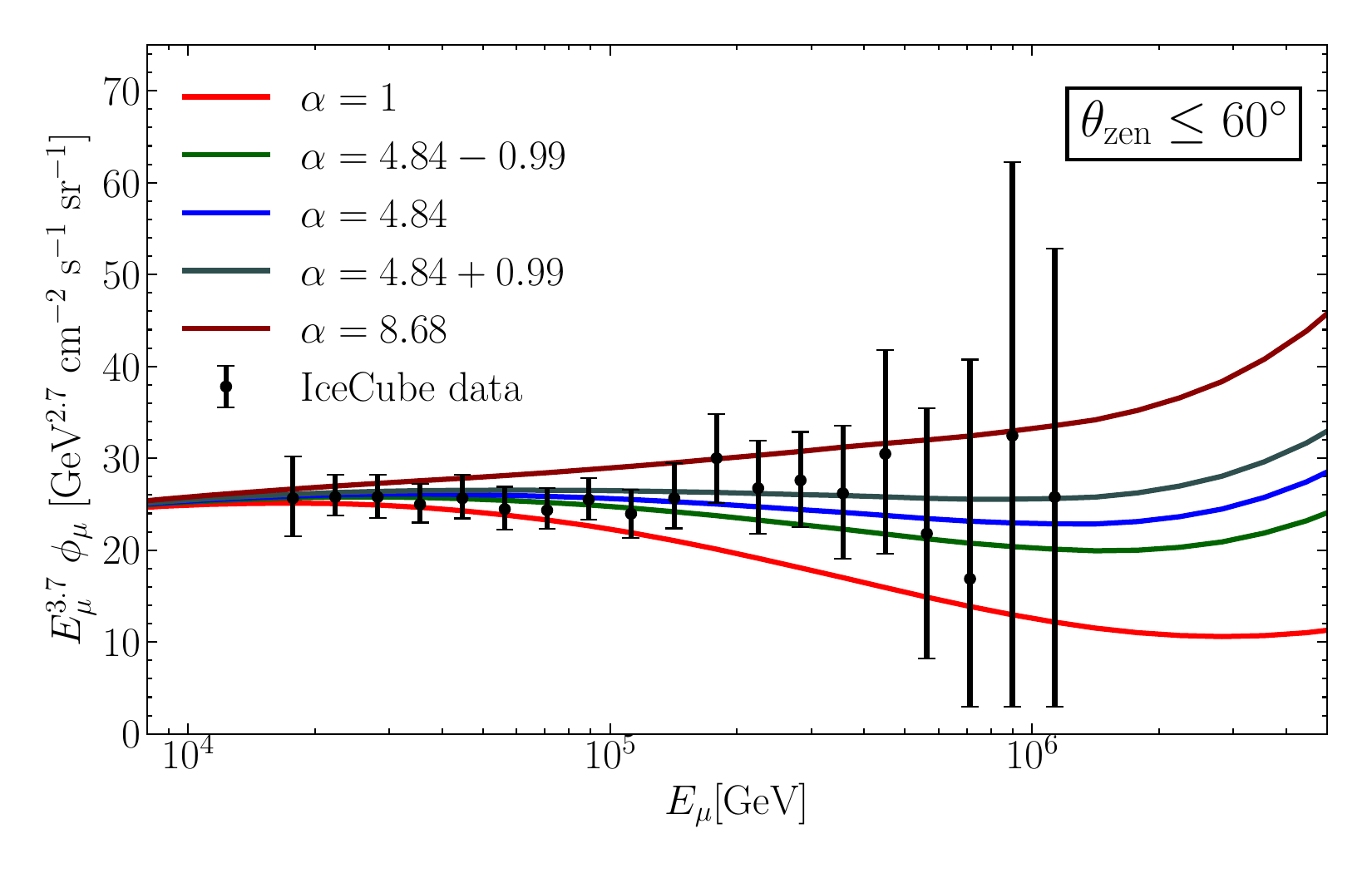}
    \hfil
    \begin{minipage}{\dimexpr 0.5\linewidth-\columnsep}
        \captionsetup{singlelinecheck=off, skip=0pt}
      \caption{Total atmospheric muon flux scaled by $E_\mu^{3.7}$ for different values of $\alpha$ compared to the scaled data from IceCube \cite{Soldin:2023lbr,Soldin:2018vak,IceCube:2015wro}. Here, $w^c_{\rm intr}=0.$}
    \label{fig:scaled-muons-alpha}
    \end{minipage}
    \vspace{-2ex}
\end{figure}

The IceCube Collaboration's atmospheric muon flux fits require a large prompt $\mu^++\mu^-$ contribution, as do our fits.
{Including a prediction for the conventional atmospheric muon flux, the IceCube Collaboration fit of the measured muon energy spectrum}
{requires a prompt contribution (combined heavy flavor prompt + unflavored prompt) of 4.75 times the ERS flux with a 90\% CL range of 2.33--9.34 times the ERS flux, } assuming the H3a cosmic-ray all-nucleon energy spectrum \cite{IceCube:2015wro}. Their angular analysis yields similar results. 
Large experimental error bars at high energies make a conclusive statement about the level of the prompt contribution difficult, as can be seen in the right panel of fig. \ref{fig:Ratio} and in 
fig. \ref{fig:scaled-muons-alpha}, where we set $w^c_{\rm intr}=0$.
Figure \ref{fig:scaled-muons-alpha} shows curves for several values of $\alpha$, including the cases $\alpha=1$, the best fit value with errors $\alpha=4.84\pm0.99$ and $\alpha=8.68$.
Forthcoming analyses of the prompt atmospheric lepton fluxes detected by IceCube and other neutrino telescopes, together with studies of unflavored meson production at accelerator experiments, are welcome, as they will contribute to resolve the tension
between IceCube experimental data on prompt atmospheric neutrino and muon fluxes.

\vskip 0.2in
\noindent
{\bf Acknowledgements}
\vskip 0.2in

This work is supported in part by the US DOE grant DE-SC-0010113 and by the DFG research unit FOR2926, project number 40824754. We are grateful to A. Fedynitch and to S. Ostapchenko for their assistance.

\bibliographystyle{JHEP-nt}
\bibliography{sample}
\end{document}